\documentstyle[preprint,aps,epsf,prd,floats]{revtex}

\def\xslash#1{{\rlap{$#1$}/}}
\def\Dsl{\hbox{/\kern-.6000em D}} %roman D
\def\vabsq#1{\left|{\bf #1}\right|^2}
\def\dsl{\,\raise.15ex\hbox{/}\mkern-13.5mu D} %this one can be subscripted
\def\bsigma{\mbox{\boldmath $\sigma$}}

\begin{document}
\preprint{\vbox{
\hbox{UCSD/PTH 97--01}
\hbox{hep-ph/9701294}
}}
\title{The HQET/NRQCD Lagrangian to order $\alpha/m^3$}
\author{Aneesh V.~Manohar}
\address{Department of Physics, University of California at San Diego,\\
9500 Gilman Drive, La Jolla, CA 92093-0319}
\date{January 1997}
\maketitle
\begin{abstract}
The HQET/NRQCD Lagrangian is computed to order $\alpha/m^3$. The computation is
performed using dimensional regularization to regulate the ultraviolet and
infrared divergences. The results are consistent
with reparametrization invariance to order $1/m^3$. Some subtleties in
the matching conditions for NRQCD are discussed. 
\end{abstract}

\section{Introduction}
Heavy quark effective theory (HQET)~\cite{iw} and non-relativistic QCD
(NRQCD)~\cite{cl,bbl} are two
effective theories that describe the interactions of almost on-shell heavy
quarks. HQET describes the interactions of quarks of mass $m$ in which the
momentum transfer $p$ is much smaller than $m$. The HQET Lagrangian has an
expansion in powers of $p/m$. HQET is typically applied to hadrons containing
a single heavy quark, such as the $B$ meson, in which $p \sim \Lambda_{\rm
QCD}$, the scale of the strong interactions. The HQET expansion is thus an
expansion in powers of $\Lambda_{\rm QCD}/m$. NRQCD describes the
interactions of non-relativistic quarks, and is typically applied to $\bar Q Q$
bound states such as the $\Upsilon$. The NRQCD Lagrangian also has an expansion
in powers of $1/m$. The momentum transfer in NRQCD is of order $m v$, so
that the small expansion parameter in NRQCD is the velocity $v$. The size of a
term in the NRQCD Lagrangian can be estimated using velocity counting
rules~\cite{bbl}. The basic difference between HQET and NRQCD can be seen from
the first two terms in the effective Lagrangian,
\begin{equation}\label{1}
{\cal L} = Q^\dagger \left(i D^0 \right) Q + \bar Q {{\bf D}^2\over 2m} Q.
\end{equation}
In HQET, the first term is of order $\Lambda_{\rm QCD}$, and the second
term is of order $\Lambda_{\rm QCD}^2/m$, whereas in NRQCD both terms are of
order $m v^2$. As a result, the quark propagator in  HQET is $i/( k^0+i \epsilon
)$, and in NRQCD it is 
\begin{equation}\label{2}
{i\over(k^0-{\bf k}^2/2m+i\epsilon)}.
\end{equation}

The HQET/NRQCD Lagrangian is computed in this paper to one loop and order
$1/m^3$. Only the terms bilinear in fermions are considered here. There are
also four-quark operators in the effective Lagrangian. Their coefficients are
order $\alpha_s$, and can be obtained simply from tree-level matching.

\section{Matching Conditions and Power Counting}
The HQET effective theory matching computation is a straightforward 
generalization of
known results to order $1/m^2$\cite{eh,flg,cbto,cb,blok}. One can compute diagrams in the full
and effective theories, and match to a given order in $1/m$. Since the HQET
propagator is $m$ independent, the HQET power counting is manifest --- one
counts powers of $1/m$ directly from the vertex factors. This means that graphs
with a vertex
of order $1/m^r$ do not make any contributions to terms of order $1/m^s$, with
$s<r$ to any order in the loop expansion.

The use of NRQCD as an effective field theory is more subtle.
NRQCD with the propagator Eq.~(\ref{2}) cannot be used as an
effective Lagrangian to compute matching corrections, since the velocity power
counting breaks down.\footnote{I would like to thank M.~Luke for extensive
discussions on this point. Some of these issues will be discussed in a future
publication. See also \cite{lm}.} The matching conditions for NRQCD should be
computed using the HQET power counting, by expanding in $p^\mu/m$. After the
HQET Lagrangian has been computed, it can be used for computing bound state
properties using the NRQCD velocity power counting rules. 
In other words, the NRQCD propagator Eq.~(\ref{2}) should be thought of as the 
infinite series
\begin{equation}\label{a3}
{1\over(k^0-{\bf k}^2/2m+i\epsilon)} = {1\over k^0} + {{\bf k}^2 \over 2 m
\left( k^0 \right)^2} + \ldots
\end{equation} 
where one uses the right hand side inside any ultraviolet divergent Feynman
graph. This is necessary when a cutoff such as dimensional regularization is
used to regulate the Feynman graphs. NRQED matching conditions have previously
been computed using a momentum space cutoff~\cite{kn}. In this case, there is
no difference between using the left or right hand sides of Eq.~(\ref{a3}).
However, a momentum space cutoff can not be used in NRQCD, since it breaks
gauge invariance.

The difference between using the two forms of Eq.~(\ref{a3}) in a loop graph can
be illustrated by a simple example. Consider the integral
\begin{equation}\label{a4}
\int_0^\infty dk^2 \ {\left(k^2\right)^a\over \left(k^2+m_1^2 \right) 
\left(k^2+m_2^2 \right)} = {\pi \over
\sin \pi a} {\left(m_1^2\right)^a - \left(m_2^2\right)^a \over m_1^2-m_2^2}.
\end{equation}
A typical NRQCD loop integral has the form Eq.~(\ref{a4}). The power
$a$ increases as one considers more and more
divergent loop graphs in the effective
theory. The denominator of a typical NRQCD loop graph has poles at $k^2\sim p^2$,
where $p$ is the external momentum, {\it and} at $k^2 \sim m^2$ where $m$ is
the quark mass. Thus the scales $m_1$ and $m_2$ in Eq.~(\ref{a4}) can be taken
to be of order $p$ and $m$, respectively. One can see immediately that the
NRQCD power counting breaks down. Loop graphs with insertions of higher
dimension operators are divergent, and can be
proportional to positive powers of $m$ because of the $\left(m_2^2\right)^a$
term in Eq.~(\ref{a4}). The positive powers of $m$ from the loop integral can
compensate for inverse powers of $m$ in the coefficient, and the
entire effective Lagrangian expansion breaks down. Now consider the same
integral, but first expand
$$
{1\over k^2 + m_2^2} = {1\over k^2} - {m^2_2\over k^4} + \ldots,
$$
evaluate the integral, and then resum the series. The answer is
\begin{equation}\label{a5}
\int_0^\infty dk^2 \ {\left(k^2\right)^a\over \left(k^2+m_1^2 \right) 
\left(k^2+m_2^2 \right)} = {\pi \over
\sin \pi a} {\left(m_1^2\right)^a \over m_1^2-m_2^2}.
\end{equation}
The integral is missing the $\left(m_2^2\right)^a$ term since it is
non-analytic at the origin for $a\not={\rm integer}$, which is where the
integral is evaluated in $4-\epsilon$ dimensions.
Equation~(\ref{a5}) only has
inverse powers of the high momentum scale $m_2\sim m$, and leads to an
acceptable effective field theory.
Thus NRQCD and HQET matching conditions are
computed in the same way, and the two Lagrangians are the same.

\section{The Lagrangian}
The continuum
NRQED effective Lagrangian at one-loop has previously been
computed~\cite{kn} using a photon mass to regulated the infrared divergences,
and a momentum space cutoff. This procedure cannot be used in a non-abelian
gauge theory such as QCD. The kinetic terms in the NRQCD Lagrangian at 
one-loop have previously been
computed by Morningstar~\cite{colin} using a lattice regulator.
The computations in this letter will be done in the continuum using
dimensional regularization for the infrared and ultraviolet divergences,
and for on-shell external states. This has the advantage
that one can freely use the equations of motion to reduce the number of
operators in the effective Lagrangian~\cite{eom}. The most general
effective Lagrangian to order $1/m^3$ (up to field redefinitions) is
\begin{eqnarray}\label{4}
&&{\cal L} = Q^\dagger \Biggl\{ i D^0 + c_2\, {{\bf D}^2\over 2 m} +
c_4\, {{\bf D}^4\over
8 m^3} + c_F\, g {{\bf \bsigma \cdot B} \over 2 m} 
 + c_D\, g {\left[{\bf D \cdot E }\right] \over 8 m^2} \nonumber \\
&& + i c_S\, g {{\bf \bsigma \cdot \left(D \times E -E \times D\right) }\over 8 m^2}
 + c_{W1}\, g { \left\{ {\bf D^2,\bsigma
\cdot B }\right\}\over 8 m^3} - c_{W2}\, g { {\bf D}^i\, {\bf \bsigma
\cdot B} \, {\bf D}^i  \over 4 m^3} \\ 
&& + c_{p'p}\, g { {\bf \bsigma \cdot D\, B \cdot D + D
\cdot B\, \bsigma \cdot D}\over 8 m^3} 
+ i c_M\, g { {\bf D\cdot \left[D \times B
\right] + \left[D \times B \right]\cdot D} \over 8 m^3}\nonumber\\
&&+ c_{A1}\, {g^2}\, {{\bf B}^2-{\bf E}^2 \over 8 m^3}- 
c_{A2}\, {g^2}\, {{\bf E}^2 \over 16 m^3} 
+ c_{A3}\, {g^2}\, {\rm Tr}\left({{\bf B}^2-{\bf E}^2 \over 8 m^3}\right)- 
c_{A4}\, {g^2}\, {\rm Tr}\left({{\bf E}^2 \over 16 m^3}\right) \\
&&+ i c_{B1}\, g^2\, { {\bf \bsigma \cdot \left(B
\times B - E \times E \right)} \over 8 m^3}
- i c_{B2}\, g^2\, { {\bf \bsigma \cdot \left( E \times E \right)} \over 8 m^3}
\Biggr\} Q ,\nonumber
\end{eqnarray}
which
is the HQET/NRQCD Lagrangian in
the special frame $v=\left(1,0,0,0\right)$, and
the notation of~\cite{kn} has been used. The covariant derivative is $D^\mu =
\partial^\mu + i g A^{\mu a} T^a = \left(D^0,-{\bf D}\right)$.
Covariant derivatives in square
brackets act only on the fields within the brackets. The other covariant
derivatives act on all fields to the right. The subscripts $F$, $S$ and $D$ stand
for Fermi, spin-orbit, and Darwin, respectively. The last seven terms in
Eq.~(\ref{4}) are not given in Ref.~\cite{kn}, since they were not
required for the computation done there. The last four terms can be omitted for
QED.

In an arbitrary frame, Eq.~(\ref{4}) can be written as
\begin{eqnarray}\label{4v}
&&{\cal L}_v = \bar Q_v \Biggl\{ 
i D\cdot v - c_2\, {D_\perp^2\over 2 m} +
c_4\, {D_\perp^4\over
8 m^3} - c_F\, g {\sigma_{\alpha\beta} G^{\alpha\beta} \over 4 m} 
- c_D\, g {v^{\alpha}\left[D^\beta_\perp G_{\alpha\beta}\right]\over 8 m^2} \nonumber \\
&& + i c_S\, g {v_\lambda
\sigma_{\alpha\beta}\left\{D^\alpha_\perp , G^{\lambda\beta}\right\}
\over 8 m^2}
 + c_{W1}\, g { \left\{ D_\perp^2,\sigma_{\alpha\beta}G^{\alpha\beta}
\right\}\over 16 m^3} - c_{W2}\, g { D^\lambda_\perp \, 
\sigma_{\alpha\beta}G^{\alpha\beta}\, D_{\perp \lambda } \over 8 m^3} \\ 
&& + c_{p'p}\, g { \sigma^{\alpha\beta}\left( D_\perp^\lambda
G_{\lambda\alpha}D_{\perp\beta} + D_{\perp\beta}
G_{\lambda\alpha}D_{\perp}^\lambda - D_\perp^\lambda
G_{\alpha\beta}D_{\perp\lambda}
\right)\over 8 m^3} 
- i c_M\, g { 
D_{\perp\alpha}\left[ D_{\perp\beta} G^{\alpha\beta}\right]+
\left[D_{\perp\beta} G^{\alpha\beta}\right]D_{\perp\alpha}
\over 8 m^3} \nonumber \\
&& + c_{A1}\, g^2\, {G_{\alpha\beta} G^{\alpha\beta} \over 16 m^3}
+ c_{A2} \, g^2\, {G_{\mu\alpha} G^{\mu\beta}v_\alpha v_\beta \over 16
m^3}+ c_{A3}\, g^2\, {\rm Tr}\left({G_{\alpha\beta} G^{\alpha\beta} 
\over 16 m^3} \right)+ c_{A4} \, g^2\, {\rm Tr}\left(
{G_{\mu\alpha} G^{\mu\beta}v_\alpha v_\beta \over 16
m^3} \right)\nonumber \\
&& -  
i c_{B1} \,g^2\, {\sigma_{\alpha\beta}\left[G^{\mu\alpha}, G_\mu{}^\beta\right] \over
16m^3} - i c_{B2} \,g^2\, {\sigma_{\alpha\beta}\left[G^{\mu\alpha}, 
G^{\nu\beta}\right] v_\mu v_\nu \over 16m^3}
\Biggr\} Q_v ,\nonumber
\end{eqnarray}
where
\begin{equation}
D_\perp^\mu = D^\mu - v^\mu\ v \cdot D.
\end{equation}
The tree level matching conditions can be obtained by integrating out the
antiquark components, and making a field redefinition to eliminate terms with
$v \cdot D$ acting on the quark fields. The ``standard'' form of the HQET
Lagrangian after integrating out the antiquark fields is
\begin{eqnarray}\label{300}
{\cal L}_v &=& \bar Q_v \left\{ i v \cdot D + i {\dsl}_\perp {1\over 2 m + i
v \cdot D} i {\dsl}_\perp \right\} Q_v \\
&=& \bar Q_v \left\{ i v \cdot D -{1\over 2 m} {\dsl}_\perp 
{\dsl}_\perp + {1\over 4 m^2} {\dsl}_\perp \left(iv\cdot D\right)
{\dsl}_\perp - {1\over 8 m^3} {\dsl}_\perp \left(iv\cdot D\right)^2
{\dsl}_\perp\right\} Q_v \nonumber
\end{eqnarray}
The field redefinition
\begin{eqnarray}\label{301}
Q_v &\rightarrow& \Bigl[ 1 - {D_\perp^2\over 8 m^2} -
{ g\sigma_{\alpha\beta} G^{\alpha\beta} \over 16m^2}
 + { D_\perp^\alpha \left(i v \cdot D\right)
D_{\perp\alpha}\over 16m^3 } 
+ {g v_\lambda D_{\perp\alpha} G^{\alpha\lambda}
\over 16m^3} \\
 && - i{ \sigma_{\alpha\beta} D_\perp^\alpha \left( i v \cdot D \right)
D_\perp^\beta \over 16m^3} 
- i {g v_\lambda \sigma_{\alpha\beta}
 D_\perp^\alpha G^{\beta\lambda} \over 16m^3} \Bigr] Q_v \nonumber
\end{eqnarray}
(where the $\sigma$ matrices are understood to be $P_v \sigma P_v$)
can be used to eliminate the time derivative terms, and put the Lagrangian into
the ``NRQCD'' form.
The result is Eq.~(\ref{4v}) with
$c_2=c_4=c_F=c_D=c_S=c_{W1}=c_{A1}=c_{B1}=1$, and
$c_{W2}=c_{p'p}=c_M=c_{A2}=c_{A3}=c_{A4}=c_{B2}=0$. The $c_A$
and $c_B$ terms are quadratic in the field strengths, and are order $g^2$. The
one loop corrections to these terms will not be computed here.

\section{Quark Form Factors and Matching Conditions}
\begin{figure}
\epsfxsize=6cm
\hfil\epsfbox{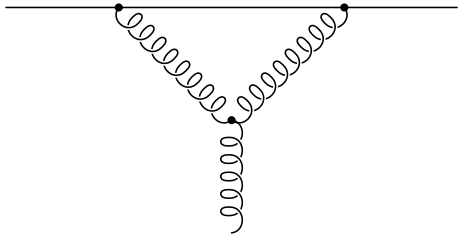}\hfill
\caption{Non-abelian contribution to the one-loop vertex correction.
\label{fig:gvertex}}
\end{figure}

A loop diagram in QCD is a function $F(\{p\},m,\mu,\epsilon)$ where $\{p\}$ are the
external momenta, $m$ is the quark mass, $\mu$ is the scale parameter of
dimensional regularization,and the computation is done in $d=4-\epsilon$
dimensions. As an example, consider the diagram~Fig.~\ref{fig:gvertex},
which gives radiative corrections to the form factors $F_1\left(q^2\right)$
and $F_2\left(q^2\right)$. In dimensional regularization, the diagram gives
the $F_1$ and $F_2$ form factors as functions of the form
$F_{1,2}(q^2/m^2,\mu/m,\epsilon)$. The form factor can be expanded as a power series in
$q^2/m^2$ at fixed $\epsilon$, followed by the limit $\epsilon\rightarrow 0$,
\begin{eqnarray}\label{3}
F_{1} &=& F_{1}\left(0\right)\left[{A_0\over \epsilon_{UV}} + 
{B_0\over \epsilon_{IR}}
+ \left(A_0+B_0\right) \log{\mu\over m} +
D_0\right]\\
&&\qquad + q^2 {dF_{1}\over d\,q^2}
\left(0\right)\left[{A_1\over \epsilon_{UV}} + {B_1\over \epsilon_{IR}}
+ \left(A_1+B_1\right) \log{\mu\over m} + D_1\right] + \ldots ,\nonumber
\end{eqnarray}
and similarly for the $F_2$ form factor.
It is conventional to label $\epsilon$ as either $\epsilon_{UV}$ or
$\epsilon_{IR}$, depending on whether the integral is ultraviolet or infrared
divergent. Ultraviolet divergences are cancelled by renormalization
counterterms. Infrared divergences cancel when a physically measurable process
is computed. Expanding the form factor in $q^2$ and then taking the limit 
$\epsilon\rightarrow 0$ gives an
expression that is analytic in $q^2$, and misses terms which are non-analytic in
$q^2$. The non-analytic terms are not needed for the calculation of the
coefficients in the effective theory, since the effective Lagrangian is analytic in
momentum. The coefficients of the effective Lagrangian are determined by
computing (for example) the difference of $F_1$ in the full theory and
effective theories. The non-analytic terms in $F_1$ cancel in the
difference, and the analytic terms determine the unknown
parameters $c_{F}\ldots c_{B2}$ in the effective Lagrangian.

Loop diagrams in HQET are functions $F(\{p\},\mu,\epsilon)$ 
times powers of the
coefficients $c_i$ in the effective Lagrangian, where $\{p\}$ are the external
momenta. All on-shell loop graphs vanish when
expanded in powers of $p$, followed by $\epsilon\rightarrow 0$. This is because
the coefficient of any power of $p$ is a dimensionally regulated integral of
the form
\begin{equation}\label{5}
\int {d^d\,k\over \left(2\pi\right)^d} f\left(k^2,k\cdot v\right).
\end{equation}
There is no dimensionful parameter in the integrand, so the integral vanishes.
The matching condition is then trivial: one takes Eq.~(\ref{3}), and throws
away the $1/\epsilon$ terms to obtain the difference of the graph in the full
and effective theory. All the $1/\epsilon$ terms in the difference are 
ultraviolet divergences (which are cancelled by renormalization counterterms),
since there are no infrared divergences in matching conditions. To see this more
explicitly, one can evaluate integrals such as Eq.~(\ref{5}) by breaking them
up into the sum of two integrals, one only ultraviolet divergent,
and the other only infrared divergent. For example,
\begin{equation}\label{11}
\int {d^d\,k\over \left(2\pi\right)^d} {1\over k^4} = 
\int {d^d\,k\over \left(2\pi\right)^d} \left[{1\over k^2\left(k^2+m^2\right)}
+{m^2\over k^4\left(k^2 + m^2\right)} \right] = {1\over 16\pi^2}\left[{1\over
\epsilon_{UV}}-{1\over \epsilon_{IR}}\right] =0,
\end{equation}
since $\epsilon_{UV}=\epsilon_{IR}=\epsilon$. A given quantity in the effective
theory is of the form
\begin{equation}\label{12}
A_{\rm eff} \left({1\over \epsilon_{UV}} - {1\over \epsilon_{IR}}\right).
\end{equation}
There can be no finite parts (the analogs of the $\left(A+B\right)\log \mu/m$ and $D$ terms in
Eq.~(\ref{3})), since the net integral is zero. A typical matching condition is
of the form
\begin{equation}\label{13}
{\rm graphs\ in\ full\ theory} = {\rm graphs\ in\ effective\ theory} + c_i,
\end{equation}
where $c_i$ is a coefficient in the effective Lagrangian. Using Eq.~(\ref{3})
and Eq.~(\ref{12}), the matching condition can be written as
\begin{equation}\label{14}
{B\over \epsilon_{IR}}
+ \left(A+B\right) \log{\mu\over m} + D = 
- A_{\rm eff} {1\over \epsilon_{IR}} + c_i.
\end{equation}
The $1/\epsilon_{UV}$ terms are cancelled by the renormalization counterterms
in the full and effective theories, respectively. The coefficients in the
effective Lagrangian have no infrared divergences. Thus $B=-A_{\rm eff}$,
and
\begin{equation}\label{15}
c_i = \left(A+B\right) \log{\mu\over m} + D,
\end{equation}
i.e. $c_i$ is obtained from Eq.~(\ref{3}) by keeping the finite pieces, and
omitting the $1/\epsilon_{UV}$ and $1/\epsilon_{IR}$ terms.

\begin{figure}
\epsfxsize=6cm
\hfil\epsfbox{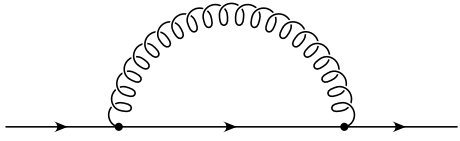}\hfill
\caption{One-loop wavefunction renormalization.
\label{fig:wave}}
\end{figure}

\begin{figure}
\epsfxsize=6cm
\hfil\epsfbox{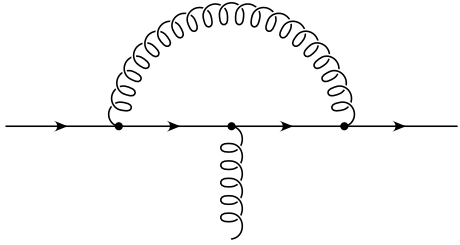}\hfill
\caption{Abelian one-loop vertex correction.
\label{fig:vertex}}
\end{figure}

The coefficients of the $Q^\dagger\left({{\bf D}^2/2 m}\right)Q$ and  
$Q^\dagger\left({{\bf D}^4
/ 8 m^3}\right)Q$ are fixed by the dispersion relation $E^2=p^2+m^2$ of QCD,
$c_2=c_4=1$. The
other terms, which all contain at least one power of the gauge field $A^\mu$,
are obtained by computing the one-loop $Q^\dagger Q A$ on-shell scattering
amplitude. The wave-function renormalization graph~Fig.~\ref{fig:wave}
and the 
vertex correction~Fig.~\ref{fig:vertex}
can be found in many textbooks on quantum
field theory~\cite{pokorski}.  In dimensional regularization, one finds that
the wave function graph is
\begin{eqnarray}\label{6}
-i \Sigma\left(p\right) &=& -i C_2\left(Q\right){\alpha_s\over 4 \pi}
\left(A\left(p^2\right)m + B\left(p^2\right) 
\xslash{p}\right) \\
A\left(p^2\right) &=& \int_0^1 dx\ \Gamma\left(\epsilon/2\right) 
\left(4-\epsilon\right) \left[ m^2 x - p^2 x (1-x) \right]^{-\epsilon/2}\\
B\left(p^2\right) &=& -\int_0^1 dx\ \Gamma\left(\epsilon/2\right) 
\left(2-\epsilon\right)\left(1-x\right) \left[ m^2 x - p^2 x (1-x) 
\right]^{-\epsilon/2}
\end{eqnarray}
where
$$
C_2\left(Q\right) = T^a T^a = {4\over 3}
$$
is the Casimir of the quark representation. The on-shell wavefunction
renormalization correction is
\begin{eqnarray}\label{7}
\delta Z &=&  - C_2\left(Q\right){\alpha_s\over 4 \pi} \left[B\left(m^2\right)
+ 2m^2 \left(
{\partial A \over \partial p^2} + {\partial B \over \partial
p^2}\right)_{p^2=m^2}\right] \nonumber \\
&=& C_2\left(Q\right) {\alpha_s\over \pi}\left[ {1\over 2 \epsilon_{UV}}
+{1\over \epsilon_{IR}} + 1 - {3\over 2} \log {m\over \mu} \right].
\end{eqnarray}

The on-shell vertex correction~Fig.~\ref{fig:vertex} can be expressed in terms of
the form-factors $F_1$ and $F_2$,
\begin{equation}\label{8}
-igT^a\ \bar u \left(p^\prime\right) \left[ F_1\left(q^2\right) \gamma^\mu +
iF_2\left(q^2\right){\sigma^{\mu \nu} q_\nu\over 2 m} \right]u\left(p\right),
\end{equation}
where $q=p^\prime-p$. The form factors are
\begin{eqnarray}\label{16}
F_1^{(V)}\left(q^2\right) &=& {\alpha\over 2\pi}\left(C_2\left(Q\right)-{1\over2}
C_2\left(ad\right)\right)\Biggl[{1\over \epsilon_{UV}} + {1\over
\epsilon_{IR}}\left(2 - {q^2\over m^2}\right) I\left(q^2/m^2\right) \nonumber \\
&& + \left(3 - {q^2\over m^2}
\right) I\left(q^2/m^2\right) -{1\over 2} J\left(q^2/m^2\right) - 
\left(1 - {q^2\over 2m^2} \right) K\left(q^2/m^2\right) -1 \Biggr]
\end{eqnarray}
and
\begin{equation}\label{10}
F_2^{(V)}\left(q^2\right) = {\alpha\over 2\pi}\left(C_2\left(Q\right)-{1\over2}
C_2\left(ad\right)\right) I\left(q^2/m^2\right).
\end{equation}
where
\begin{equation}\label{19}
I\left(q^2/m^2\right) = \int_0^1 dx\ {m^2\over m^2 - q^2 x (1-x)}
\end{equation}
\begin{equation}\label{18}
J\left(q^2/m^2\right) = \int_0^1 dx\ \log {m^2 - q^2 x (1-x) \over
\mu^2}
\end{equation}
\begin{equation}\label{17}
K\left(q^2/m^2\right) = \int_0^1 dx\ {m^2\over m^2 - q^2 x (1-x)} \log {m^2 - 
q^2 x (1-x)\over
\mu^2},
\end{equation}
and 
$$C_2\left(ad\right)=3$$
is the Casimir of the adjoint representation.
Expanding to order $q^2/m^2$ gives
\begin{eqnarray}\label{20}
F_1^{(V)} &=& {\alpha\over \pi} \left(C_2\left(Q\right)-{1\over2}
C_2\left(ad\right)\right)\left[{1\over 2 \epsilon_{UV}} + {1\over
\epsilon_{IR}} + 1 - {3\over 2} \log{m\over\mu} + {q^2\over m^2} \left(
-{1\over 3}{1\over \epsilon_{IR}} - {1\over 8} + {1\over 3} \log{m\over\mu}
\right) \right], \\
F_2^{(V)} &=& {\alpha\over \pi} \left(C_2\left(Q\right)-{1\over2}
C_2\left(ad\right)\right)\left[{1\over 2} + {q^2\over 12 m^2} \right].
\end{eqnarray}

The final diagram is the non-abelian vertex correction~Fig.~\ref{fig:gvertex}. This
is computed in background field Feynman gauge, which preserves gauge
invariance. The resulting diagram can also be evaluated in terms of the $F_1$
and $F_2$ form-factors,
\begin{eqnarray}\label{22a}
F_1^{(g)} &=& {\alpha_s\over 8 \pi} C_2\left(ad \right)
\int_0^1 dx\int_0^{1-x} dy \Biggl\{
-\Gamma\left(1+\epsilon/2\right)\Bigl[ 2 q^2 \left(x+y\right)\nonumber  \\
&& + 2 m^2 \left(1-x-y\right) \left(2 \left(x+y\right) +
\left(4-\epsilon\right) \left(1-x-y\right)\right)\Bigr]  
\left(m^2
\left(x+y-1\right)^2 - q^2 x y \right)^{-1-\epsilon/2} \nonumber \\
&&\qquad + \left(2-\epsilon\right) \Gamma\left(\epsilon/2\right)
\left(m^2
\left(x+y-1\right)^2 - q^2 x y \right)^{-\epsilon/2}\Biggr\} \nonumber \\
&=& {\alpha_s\over 8 \pi} C_2\left(ad \right)\left[{2\over \epsilon_{UV}}
+{4\over \epsilon_{IR}} + 4 -
6 \log{m\over\mu} + {q^2\over m^2}\left(-{3\over \epsilon_{IR}} -1 + 3 
\log{m\over \mu} \right) + \ldots \right],
\end{eqnarray}
\begin{eqnarray}\label{21}
F_2^{(g)} &=& - {\alpha_s\over 4 \pi} C_2\left(ad \right) m^2 
\Gamma\left(1+\epsilon/2\right)
\int_0^1 dx\int_0^{1-x} dy \left(1-x-y\right)\nonumber  \\
&& \qquad \times \left(\epsilon+\left(2-\epsilon\right)\left(x+y\right)\right) \left(m^2
\left(x+y-1\right)^2 - q^2 x y \right)^{-1-\epsilon/2} \nonumber \\
&=& {\alpha_s\over 8 \pi} C_2\left(ad \right)\left[{4\over \epsilon_{IR}} + 6 -
4 \log{m\over\mu} + {q^2\over m^2}\left({4\over \epsilon_{IR}} + 1 - 4
\log{m\over \mu} \right) + \ldots \right].
\end{eqnarray}

The total on-shell form factors at one loop are given by
\begin{eqnarray}\label{22}
F_1 &=& 1 - \delta Z + F_1^{(V)} + F_1^{(g)} \\
&=& 1 + {\alpha_s\over \pi}{q^2\over m^2}
\left[ \left( -{1\over 3 \epsilon_{IR}} -{1\over 8} +{1\over 3}
\log {m \over \mu} \right)C_2\left(Q\right)+ \left( -{5\over 24 \epsilon_{IR}} 
-{1\over 16} +{5\over 24} \log {m \over \mu} \right)C_2\left(ad\right)\right],
\nonumber \\
F_2 &=& F_2^{(V)} + F_2^{(g)} \label{23}\\
&=&  {\alpha_s\over \pi}
\left[ {1\over 2} C_2\left(Q\right) +  
\left( {1\over 2 \epsilon_{IR}} + {1\over 2}  - {1\over 2} \log {m \over \mu}
\right)C_2\left(Q\right)\right] \nonumber \\
&&\qquad + 
{\alpha_s\over \pi}{q^2\over m^2}
\left[{1\over 12} C_2\left(Q\right) + 
\left( {1\over 2 \epsilon_{IR}}
+ {1\over 12} - {1 \over 2} \log {m \over \mu} \right)
C_2\left(ad\right)\right]. \nonumber
\end{eqnarray}
The total form-factor $F_1\left(0\right)$ is unity, since gauge invariance is
preserved by the background field method.

The scattering amplitude for a low-momentum heavy quark off a background 
vector potential can be computed by expanding Eq.~(\ref{8}), and multiplying by
$\sqrt{m/E}$ for the incoming and outgoing quarks. If ${\bf p}$ is
the three-momentum of the incoming quark, ${\bf p}^\prime$ is the
three-momentum of the outgoing quark, and ${\bf q = p^\prime -p}$,
one finds that the effective interaction is
\begin{equation}\label{23-1}
-i g T^a\, u^\dagger_{NR} \left( {\bf p}^\prime \right)\left[ A^{0a}j^0 -
{\bf A}^a \cdot {\bf j} \right] u_{NR} \left({\bf p}\right),
\end{equation}
where
\begin{equation}\label{23-2}
j^0 =  F_1\left(q^2\right) 
\left\{  1 -  {1\over 8 m^2}\vabsq{q} + {i\over 4 m^2} 
{\bf \bsigma \cdot \left(p^\prime \times p\right)} \right\} 
+  F_2\left(q^2\right) \left\{  - {1\over 4 m^2}\vabsq{q} + {i\over 2 m^2} 
{\bf \bsigma \cdot \left(p^\prime \times p\right)} \right\}, 
\end{equation}
and
\begin{eqnarray}\label{23-3}
{\bf j} &=& F_1\left(q^2\right)\Biggl\{
{1\over 2 m}\left({\bf p + p^\prime}\right)  + {i\over 2m} {\bf \bsigma
\times q } - {i\over 8 m^3} \left( \vabsq p + \vabsq{p^\prime} \right)
{\bf \bsigma \times  q } \\
&& -{i\over 16 m^3} \left( \vabsq{p^\prime} - \vabsq p \right)
{\bf \bsigma \times\left( p + p^\prime\right)}
- {1\over 8 m^3}\left(\vabsq{p^\prime}+ \vabsq p \right) 
{\bf \left( p + p^\prime\right)} - {1\over  16 m^3}
\left(\vabsq{p^\prime} - \vabsq p \right) {\bf q}
\Biggr\} \nonumber \\
&&+ F_2\left(q^2\right)\Biggl\{
{i\over 2m} {\bf \bsigma
\times q } - {i\over 16 m^3} \vabsq q
{\bf \bsigma \times q } 
- { 1\over 16 m^3} \vabsq q { \bf \left( 
p^\prime + p \right) } \nonumber \\
&& -  { 1\over 16 m^3}\left(\vabsq{p^\prime} - \vabsq p\right)
{\bf q }  -{i\over 8 m^3} \left( \vabsq{p^\prime} - \vabsq p \right)
{\bf \bsigma \times \left( p^\prime + p \right)} 
+ {i \over 8 m^3}
{\bf \bsigma \cdot \left( p^\prime + p \right)} \left({\bf p^\prime \times p}
\right) \Biggr\}.\nonumber
\end{eqnarray}

Comparing Eqs.~(\ref{23-2},\ref{23-3}) with the scattering amplitude in the effective theory
from the Lagrangian Eq.~(\ref{4}) gives
\begin{eqnarray} \label{25}
c_F &=& F_1 + F_2 \\
c_D &=& F_1 + 2 F_2 + 8 F_1^\prime \\
c_S &=& F_1 + 2 F_2 \\
c_{W1} &=& F_1 + {1\over 2 } F_2 + 4 F_1^\prime + 4 F_2^\prime \\
c_{W2} &=& {1\over 2 } F_2 + 4 F_1^\prime + 4 F_2^\prime \\
c_{p^\prime p} &=& F_2 \\
c_M &=& -{1\over 2} F_2 - 4 F_1^\prime
\end{eqnarray}
where
\begin{equation}
F_i \equiv F_i \left(0\right), \qquad F_i^\prime \equiv \left.{d F_i \over
d\left(q^2/m^2\right)}\right|_{q^2=0}
\end{equation}
Note that the nine parameters (including $c_2$ and $c_4$) 
in the effective Lagrangian Eq.~(\ref{4}) are determined in terms of only three
independent constants, $F_2$, $F_1^\prime$, and $F_2^\prime$, since $F_1=1$.
Reparametrization invariance~\cite{reparam} gives six linear relations among
the coefficients. This will be discussed in more detail in the next section.

The explicit expressions for the coefficients are obtained using
Eqs.~(\ref{22},\ref{23}):
\begin{eqnarray} \label{26}
c_F &=& 1+{\alpha\over \pi}\left[ {1\over 2}
C_2\left(Q\right) + \left( {1\over 2} -
{1\over 2}\log {
m \over  \mu} \right)C_2\left(ad\right)\right]  \\
c_D &=& 1+{\alpha\over \pi}\left[ \left( 
{8\over 3} \log {m \over \mu} \right)C_2\left(Q\right) + \left( {1\over 2} +
{2\over 3}\log {
m \over  \mu} \right)C_2\left(ad\right)\right]  \\
c_S &=& 1+{\alpha\over \pi}\left[ C_2\left(Q\right)+ 
\left( 1 -
\log {m \over  \mu} \right)C_2\left(ad\right)\right]  \\
c_{W1} &=& 1+{\alpha\over \pi}\left[ \left( {1\over 12} +
{4\over 3} \log {m \over \mu} \right)C_2\left(Q\right) + \left( {1\over 3} -
{17\over 12}\log {
m \over  \mu} \right)C_2\left(ad\right)\right] \\
c_{W2} &=& {\alpha\over \pi}\left[ \left( {1\over 12} +
{4\over 3} \log {m \over \mu} \right) C_2\left(Q\right)+ \left( {1\over 3} -
{17\over 12}\log {
m \over  \mu} \right)C_2\left(ad\right)\right] \\
c_{p^\prime p} &=& {\alpha\over \pi}\left[ {1\over 2} C_2\left(Q\right)
+ \left({1\over2} -{1\over 2}\log {
m \over  \mu} \right)C_2\left(ad\right)\right]\\
c_M &=& {\alpha\over \pi}\left[ \left( {1\over 4} -{4\over 3
} \log {m \over \mu} \right)C_2\left(Q\right) - \left( {7\over 12}\log {
m \over  \mu} \right)C_2\left(ad\right)\right]
\end{eqnarray}
The results for NRQED can be obtained by setting $C_2\left(ad\right)=0$ and
$C_2\left(Q\right)=1$, and agree with those found
in~\cite{kn}, with the replacement
\begin{equation}\label{27}
\log \mu \rightarrow \log 2 \Lambda - {5\over 6}
\end{equation}
The difference in finite parts is because the NRQED integrals were evaluated in
Ref.~\cite{kn} using a momentum space cutoff, instead of using dimensional 
regularization. The results for the $1/m$ operators agree with known results
for HQET~\cite{eh,flg}. The $1/m^2$ matching conditions at tree-level, and the
$\mu$ dependence at one-loop also agree with known results~\cite{cbto,cb,blok}.
Note that $c_F$ is independent of $\mu$ in QED. This is easy to
see if one computes the renormalization of the magnetic moment operator in the
effective theory in Coulomb gauge, in which all transverse photon interactions
are suppressed by $1/m$. 

\begin{figure}
\epsfxsize=6cm
\hfil\epsfbox{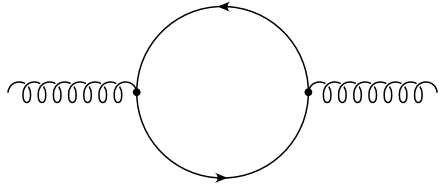}\hfill
\caption{Quark contribution to the vacuum polarization.
\label{fig:qloop}}
\end{figure}

\begin{figure}
\epsfxsize=6cm
\hfil\epsfbox{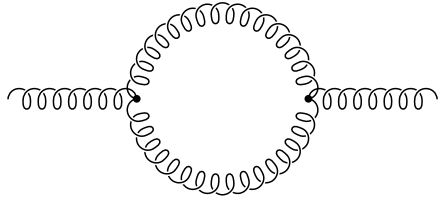}\hfill
\caption{Gluon contribution to the vacuum polarization.
\label{fig:gloop}}
\end{figure}

The discussion so far has concentrated on the fermion part of the effective
Lagrangian. There is, in addition, the pure gauge field part of the effective
action. The one-loop correction to the gluon propagator is shown in
Figs.~(\ref{fig:qloop})
and (\ref{fig:gloop}). 
The gluon diagram is the same
in QCD and in HQET, so the one-loop matching condition is from the quark
vacuum polarization diagram. This gives the effective
action~\cite{cho,shifman,balk}
\begin{equation}\label{28}
{\cal L} = - {1\over 4}d_1 G_{\mu \nu}^A G^{A\mu \nu}  
+{d_2\over m^2} G_{\mu \nu}^A D^2 G^{A\mu \nu} +{d_3\over m^2}g f_{ABC}
G_{\mu\nu}^A G_{\mu\alpha}^B G_{\nu\alpha}^C  
+{\cal O}\left({1\over m^4}\right),
\end{equation}
with 
\begin{eqnarray}
d_1 &=& 1-{ \alpha\over 3 \pi} T\left(Q\right) \log m^2/\mu^2, \nonumber \\
d_2 &=& {\alpha\over 60 \pi} T\left(Q\right),\\
d_3 &=& {13\alpha\over 360 \pi} T\left(Q\right),\nonumber
\end{eqnarray}
where 
$$T\left(Q\right)=1/2$$ 
is the index of the quark representation. The
identity
$$
0= \int 2\ D^\mu\, G_{\mu \alpha}^A\, D_\nu\, G^{\nu \alpha A} +
2\ g\, f_{ABC}\, G_{\mu\nu}^A G_{\mu\alpha}^B G_{\nu\alpha}^C + 
G_{\mu \nu}^A\, D^2\, G^{A\mu \nu}
$$
has been used to eliminate $D^\mu\, G_{\mu \alpha}^A\, D_\nu\, 
G^{\nu \alpha A}$ from Eq.~(\ref{28}). 

\section{Reparametrization Invariance}
The coefficients of operators in the HQET Lagrangian are constrained by
reparametrization invariance~\cite{reparam}. The reparametrization invariant
spinor field $\Psi_v$ is given by
\begin{equation}\label{101}
\Psi_v = \Lambda\left(w,v\right) \psi_v,
\end{equation}
where $\psi_v$ is the conventional heavy quark field that satisfies
\begin{equation}\label{102}
\xslash v \psi_v = \psi_v,
\end{equation}
$\Lambda(w,v)$ is the Lorentz transformation matrix
\begin{equation}\label{103}
\Lambda\left(w,v\right) = {1 + \xslash w\ \xslash v \over 
\sqrt{2 \left(1 + w\cdot v\right)}},
\end{equation}
and
\begin{equation}\label{104}
w^\mu = {v^\mu + i D^\mu/m \over \left|v^\mu + i D^\mu/m\right|}.
\end{equation}
One needs to choose a particular operator ordering for the covariant
derivatives; different orderings are related to each other by field
redefinitions.

It is simplest to consider the consequences of reparametrization invariance
when $D^\mu \rightarrow \partial^\mu$ in Eq.~(\ref{101}). Then there is no
operator ordering ambiguity, and the field $\Psi_v$ can be written as
\begin{eqnarray}\label{105}
\Psi_v &=& \left[ {1 \over 2 \sqrt{\left(m+i\partial\cdot v\right)^2+
\left(i{\partial_\perp}\right)^2} \left(m+i
\partial \cdot v + \sqrt{\left(m+i\partial\cdot v\right)^2+\left(i{
\partial_\perp}\right)^2} \right)}\right]^{1/2}\\
&& \times \left[ m+ +i \partial \cdot v +
i\xslash \partial_\perp + \sqrt{\left(m+i\partial\cdot v\right)^2
+\left(i{\partial_\perp}\right)^2} \right] \psi_v ,\nonumber
\end{eqnarray}
where
\begin{equation}\label{106}
\partial^\mu_\perp = \partial^\mu - v^\mu \ \partial \cdot v.
\end{equation}
If one uses Eq.~(\ref{105}), replaces $\psi_v$ by the spinor 
$u_{NR} e^{-i p \cdot x}$, with $p^2=m^2$, $\gamma^0 u_{NR} = u_{NR}$, 
$\bar u_{NR}^\dagger u_{NR}=1$,
and $v=\left(1,0,0,0)\right)$, the field $\Psi_v$
reduces to the spinor $u e^{-i p \cdot x}$, that satisfies 
the Dirac equation $\xslash p u = m u$, and is normalized so that $\bar
u u=1$. This shows that reparametrization invariance
will determine the coefficients of the $1/m$ suppressed operators which are 
fixed by relativistic invariance.

The reparametrization invariant kinetic term is 
\begin{equation}\label{107}
\bar\Psi_v i\xslash \partial\, \Psi_v = \bar\psi_v\left[ 
\sqrt{\left(m+i\partial \cdot v\right)^2+ \left(i\partial_\perp\right)^2} -m
\right] \psi_v. 
\end{equation} 
This is not the same as the terms in the
Lagrangian Eq.~(\ref{4}).  The reparametrization invariant field
Eq.~(\ref{101}) does not automatically  produce a Lagrangian in the
``standard'' NRQCD form. However, one can convert Eq.~(\ref{107}) to this form
by making a field redefinition, 
\begin{equation}\label{108} 
\psi_v = \left[
{\sqrt{\left(m+i\partial \cdot v\right)^2+ \left(i\partial_\perp\right)^2} +m
\over  m + i  \partial \cdot v + \sqrt{-\left(i\partial_\perp\right)^2+m^2}
}\right]^{1/2}\psi_v^\prime 
\end{equation} 
The kinetic energy term in the
primed field is 
\begin{equation}\label{109} \bar \psi_v^\prime \left[ m + i
\partial \cdot v - \sqrt{-\left(i\partial_\perp\right)^2+m^2}\right]
\psi_v^\prime. 
\end{equation} 
which when expanded gives Eq.~(\ref{4}) with
$c_2=c_4=1$. Thus $c_2=c_4=1$ follows from reparametrization invariance. The
transformation factor in Eq.~(\ref{108}) when applied to on-shell spinors
(instead of fields) reduces to $\sqrt{m/E}$. This is the same as the flux
factor for the incoming and outgoing particles which was included in
Eq.~(\ref{23-1}).

To determine the constraints of reparametrization invariance on the effective
Lagrangian, consider Eq.~(\ref{105}) with the gauge fields included, i.e.\ with
$\partial \rightarrow D$. Expanding to order $1/m^3$ gives
\begin{equation}\label{200}
\Psi_v = \left[1+ A + {i{\dsl}_\perp \over 2 m} B \right] \psi_v
\end{equation}
where
\begin{eqnarray}
A &=& 1 - {\left(iD_\perp\right)^2  \over 8  m^2 } 
+  {\left(i D_\perp\right)^2 \left(i v \cdot D \right) \over
4 m^3} 
\nonumber \\
B &=& 1 - {iv\cdot D\over  m} - {3\left(iD_\perp\right)^2  \over 8  m^2 }
+ {\left(iv\cdot D\right)^2 \over m^2 }.
\end{eqnarray}
A particular ordering has been chosen for the operators in Eq.~(\ref{200}). A
different ordering gives an effective Lagrangian that is related by a field
redefinition. 

The most general reparametrization invariant Lagrangian is a
linear combination of invariant terms, such as $\bar \Psi_v \left(
\xslash v + i \dsl/m \right) \Psi_v$, $\bar\Psi_v
\sigma^{\alpha\beta}G_{\alpha\beta} \Psi_v$, etc.  The effective Lagrangian
obtained in this way is not in the form Eq.~(\ref{4v}), but it can be converted
into that form by field redefinitions that preserve $\xslash v \psi_v =
\psi_v$. One finds by a straightforward (but not very enlightening computation)
that the effective Lagrangian is a linear combination of the invariant linear
combinations
\begin{eqnarray}
&i v \cdot D + O_2 + O_4 + O_F + O_D + O_S +O_{W1}, & \nonumber \\
&2 O_F + 4 O_D + 4 O_S +  O_{W1} + O_{W2} + 2 O_{p^\prime p}
-  O_{M},  &\nonumber\\
&  O_{W1} + O_{W2}, & \\
&2 O_D + O_{W1} + O_{W2} - O_M ,& \nonumber\\
&O_{A1},\ O_{A2},\ O_{A3},\ O_{A4},\ O_{B1},\ O_{B2},  & \nonumber
\end{eqnarray}
up to terms of order $1/m^4$. Here $O_F$,
etc.\ are the operator coefficients of $c_F$, etc.\ in Eq.~(\ref{4v}). The above linear
combinations imply the constraints
\begin{eqnarray}
c_2 &=& 1, \nonumber\\
c_4 &=& 1, \nonumber\\
c_S &=& 2 c_F - 1,\\
c_{W2} &=& c_{W1} - 1 ,\nonumber\\
c_{p^\prime p} &=& c_F - 1, \nonumber\\
2c_M &=& c_F - c_D  .\nonumber
\end{eqnarray}
which are satisfied by Eqs.~(\ref{25}),(\ref{26}).

\section{Conclusions}
The HQET/NRQCD Lagrangian has been computed to one loop and order $1/m^3$,
and has been shown to be reparametrization invariant to order
$\alpha/m^3$. The original form of the NRQCD propagator Eq.~(\ref{2}) cannot be
used to compute the effective Lagrangian by matching to QCD. Instead, one must
treat the propagator as an infinite series, and resum the series {\it after}
doing the loop integral. As a result, the matching computations for NRQCD and
HQET are the same.
It is
straightforward to obtain the effective Lagrangian (in the one-quark sector) to
higher orders in $1/m$ by expanding the form factors $F_{1,2}$ and the spinors
in the computation of Eqs.~(\ref{23-1}),(\ref{23-2}) to higher orders. No further
Feynman graphs need to be evaluated.

\acknowledgments
I am indebted to M.~Luke for numerous discussions about matching conditions in
non-relativistic effective field
theories. This work was supported in part by a
Department of Energy grant DOE-FG03-90ER40546,


\begin{references}

\bibitem{iw} N. Isgur and M.B. Wise, Phys. Lett. {\bf B232} 113 (1989); Phys.
Lett. {\bf B237} 527 (1990).

\bibitem{cl} W.E. Caswell and G.P. Lepage, Phys. Lett. {\bf B167}, 437 (1986).

\bibitem{bbl} G.T. Bodwin, E. Braaten, and G.P. Lepage, Phys.\ Rev.\ {\bf D51},
1125 (1995).

\bibitem{eh} E. Eichten and B. Hill, Phys.\ Lett.\ {\bf B243}, 427 (1990).

\bibitem{flg} A.F.~Falk, B.~Grinstein, and M.E.~Luke, Nucl.\ Phys.\ {\bf
B357}, 185 (1991).

\bibitem{cbto} C.~Balzereit and T.~Ohl, Phys.\ Lett.\ {\bf B386}, 335 (1996).

\bibitem{cb} C.~Bauer, Diplom Thesis, Karlsruhe 1996 (unpublished).

\bibitem{blok} B.~Blok, J.G.~Korner, D.~Pirjol, and J.C.~Rojas, hep-ph/9607233.

\bibitem{kn} T.~Kinoshita and M.~Nio, Phys.\ Rev.\ {\bf D53}, 4909 (1996).

\bibitem{colin} C.~Morningstar, Phys.\ Rev.\ {\bf D48}, 2265 (1993),
{\bf D50}, 5902 (1994).

\bibitem{eom} H.D.~Politzer, Nucl.\ Phys. {\bf B172}, 349 (1980).

\bibitem{pokorski} See for example, S.~Pokorski, {\it Gauge Field Theories},
Cambridge University Press, 1987.

\bibitem{lm} M.E.~Luke and A.V.~Manohar, hep-ph/9610534.

\bibitem{reparam} M.E.~Luke and A.V.~Manohar, Phys.\ Lett.\ {\bf B286}, 348
(1992).

\bibitem{cho} P.~Cho and E.H.~Simmons, Phys.\ Rev.\ {\bf D51}, 2360 (1995).

\bibitem{shifman} V.~Novikov, M.~Shifman, A.~Vainshtein, and V.~Zakharov,
Fort.\ Phys.\ {\bf 32}, 585 (1984).

\bibitem{balk} S.~Balk, J.G.~Korner, and D.~Pirjol, Nucl.\ Phys.\ {\bf B428},
499 (1994).


\end{references}
\end{document}